\documentclass{article}
\def\msun{M_\odot}
\def\ten#1{\times 10^{#1}}

\begin{document}
\title{{\bf Black Holes Must Die }}
\author{Neal Dalal \& Kim Griest\\
{\small\it Physics Department 0319, University of California, San Diego,
La Jolla CA 92093}}
\maketitle

\noindent
\centerline{\bf Abstract}\break\noindent
In light of recent evidence suggesting a nonzero present-day
cosmological constant \cite{perlmutter}, Adams, Mbonye, \&
Laughlin\cite{aml} have considered the evolution of black holes in the
presence of vacuum energy.  Using the assumption that $\Lambda$
remains constant with time and a conjecture based on a paper by
Mallett \cite{malletta}, they reach the remarkable conclusion that
black holes with current mass greater than $\sim 2 \ten{-9}\msun$ will
not Hawking evaporate in the distant future, but will instead absorb
vacuum energy and grow to roughly the de Sitter horizon size.  In this
letter we reexamine black hole evaporation in the presence of a
cosmological constant, and find instead that all known black holes
will eventually evaporate.

\section{Naive Estimate}
It is well known that a black hole of mass $M$
in otherwise empty spacetime radiates
as a blackbody at the Hawking temperature 
\begin{equation}
\label{eq-hawkingT}
T_H = 1 /(8\pi GM) = 6.15 \ten{-8} (\msun/M){\rm K}.
\end{equation}
It is also well known that inertial observers in a de Sitter universe
with cosmological constant $\Lambda$
feel themselves to be immersed in a bath of thermal particles at 
the Gibbons-Hawking temperature \cite{wald}
\begin{equation}\label{eq-gibbonsT}
T_{\rm GH} = {1\over 2\pi} \left({\Lambda\over 3}\right)^{1/2} \approx
2.3 \ten{-30}\left({{\Omega_V}\over{0.7}}\right)
\left({h\over{0.7}}\right)^2{\rm K}.
\end{equation}
In the above equation we have used $\Omega_V=\rho_V/\rho_{crit}$,
to relate the vacuuum energy density, $\rho_V = \Lambda/(8 \pi G)$,
to the critical energy density,
$\rho_{crit}=3 H_0^2/(8 \pi G)$,
with $\hbar=c=k_B=1$.  Current measurements
imply a Hubble constant $h = H_0/ 100 {\rm km/s/Mpc} \approx 0.7$, and 
$\Omega_V \approx 0.7$, 
giving $\rho_V \approx 3.6$ keV/cm$^3 = 2.78 \ten{-11}$ eV$^4$,
and $\Lambda \approx 4.7 \ten{-66}$ eV$^2$.

Consider now a black hole in a de~Sitter universe,
with the black hole horizon at approximately $r_S\equiv 2GM$
and cosmological de~Sitter horizon at approximately
$r_{\rm dS}\equiv\sqrt{3/\Lambda}$.
If the two horizons are sufficiently far apart, that is 
if $r_S \ll r_{\rm dS}$, then to first approximation
one might expect that the Gibbons-Hawking radiation and the 
black hole's Hawking radiation would remain essentially unchanged.
The black hole should accrete some of the Gibbons-Hawking radiation,
however, and so the net rate of mass loss should be diminished.  
Note that for any currently known black hole, accretion of the 
cosmic microwave background radiation and 
other interstellar and intergalactic matter is a much larger effect
than either of the two terms above, 
but we are considering the distant future where the de Sitter expansion
has rendered such accretion negligible.

An order of magnitude estimate of the black hole's absorption cross
section should be something like $4\pi r_S^2$, the black hole's
surface area.  Naively, then, the change in the mass of the black hole is
${\dot M}  \approx 4\pi \sigma r_H ^2 (T_{\rm GH}^4 - T_H^4)$.
Thus, there should exist some critical mass above which black holes
have a net accretion of energy, and thus never evaporate by the
Hawking process.  Setting $\dot M  = 0$ gives
$M_{\rm crit} = (1/4G) \sqrt{3/\Lambda} \approx 2.7\ten{22}\msun$ 
for current estimates of $\Lambda$.
Thus, naive considerations lead us to expect that black holes of size
comparable to the de Sitter horizon could escape evaporation, whereas
all smaller black holes should eventually evaporate.  For comparison, all
known black holes are in the mass range $1 \msun$ to $3\ten{9}\msun$,
comfortably in the evaporation regime.

\section{Two-dimensional calculation}
While the simple estimates above are suggestive, it would be better to
perform a more self-consistent calculation.  In order to find the
black hole's evaporation rate, we would like to compute the
renormalized expectation value of the stress tensor, $\langle
T_{\mu\nu}\rangle$, which encodes the flow of energy into and out of
the hole.  Unfortunately, this is quite difficult to calculate in four
spacetime dimensions.  It is far easier to calculate the renormalized
$\langle T_{\mu\nu}\rangle$ for massless scalar fields in two
dimensional spacetimes, by exploiting the conformal triviality of
these spacetimes.  For spherically symmetric geometries, one may
reduce the dimensionality of the problem to two by suppressing the
angular coordinates.  Ford and Parker \cite{ford} have shown that the
two dimensional calculation is equivalent to the full four dimensional
calculation in the geometric optics limit.

With this in mind, it is possible to compute the two-dimensional 
$\langle T_{\mu\nu}\rangle$.  We may describe the black hole embedded
in de Sitter spacetime by the Schwarzschild-de~Sitter geometry,
\begin{equation}
ds^2=-f(r)dt^2 + [f(r)]^{-1}dr^2 + r^2 d\Omega^2\\
\end{equation}
\begin{equation}
f(r)=1-{{2GM}\over r}-{{\Lambda r^2}\over 3}=
{\Lambda\over {3r}}(r-r_h)(r_c-r)(r+r_h+r_c)
\end{equation}
with black hole mass $M$ and cosmological constant $\Lambda$.
Here, $r_h$ and $r_c$ are respectively the radii of the black hole horizon and
cosmological horizon \cite{tt90}, found from 
$f(r_{\rm horizon})=0$. 
When the two horizons are far from each other $r_h \approx r_S$ and
$r_c \approx r_{\rm dS}$ defined above.
Suppressing the angular coordinates, and changing to double null
coordinates $u=t-r_\ast, v=t+r_\ast$, with $r_\ast$ the usual
Regge-Wheeler ``tortoise'' radial coordinate given by
$dr_\ast=dr/f(r)$, the line element takes the form
$ds^2=-f(r)dudv$.  

With this, we can now compute the components of the stress
tensor.  For times long after the black hole has formed, the stress
tensor becomes \cite{tt90,fordb,davies,hiscock} 
\begin{eqnarray}
T_{vv}&=&{1\over{192\pi}}\left[2f(r){{d^2f}\over{dr^2}}-
\left({{df}\over{dr}}\right)^2
+\left(\left.{{df}\over{dr}}\right|_{r=r_c}\right)^2\right]\label{tvv}\\
T_{uu}&=&{1\over{192\pi}}\left[2f(r){{d^2f}\over{dr^2}}-
\left({{df}\over{dr}}\right)^2
+\left(\left.{{df}\over{dr}}\right|_{r=r_h}\right)^2\right]\\
T_{uv}&=&{1\over{96\pi}}f(r){{d^2f}\over{dr^2}}
\end{eqnarray}
where the boundary terms are evaluated at the respective past null
infinities, which are the horizon radii $r_c,r_h$.  As $u$ is an
outgoing coordinate, $T_{uu}$ represents the outwards flow of energy,
and similarly $T_{vv}$ represents the inwards flow of energy.  Note
that with $\Lambda=0$, far from the hole $T_{uu}$ is given by the
constant term.  This is the Hawking term, representing (2D) radiation
at the Hawking temperature.  This can be seen by noting that the
surface gravity at the horizon takes the form \cite{hiscock}
\begin{equation}
\kappa={1\over 2}\left.{{df}\over{dr}}\right|_{r=r_h}
\end{equation}
and since the Hawking temperature is given by $T_H=\kappa/2\pi$, we
have 
\begin{equation}
\sigma_{2D}T_H^2={1\over{192\pi}}\left(\left.{{df}\over{dr}}
\right|_{r=r_h}\right)^2
\end{equation}
where $\sigma_{2D}=\pi/12$.  
In the absence of the cosmological term, this outward flux of positive
energy from the black hole would be precisely matched by an inward
flux of negative energy given by $T_{vv}(r_h)$ \cite{hiscock}.  
The cosmological horizon brings an extra term to the ingoing
component, representing positive flux of energy into the black hole.

As a side remark, we note that with no black hole ($M=0$), the stress
tensor reduces to $T_{uu}=T_{vv}=0,\ T_{uv}=f''f/96\pi$.  This is just
as we expect -- the stress tensor is isotropic and proportional to the
metric tensor \cite{fulling}.  
The conclusions of Adams et al. are based partially upon a calculation
by Mallett \cite{mallettb}, who left out the boundary term and therefore
did not have $T_{\mu\nu}\propto g_{\mu\nu}$ for $\Lambda=0$.  

With the above expressions for $T_{\mu\nu}$, we need only plug in our
form for $f$ to compute the stress tensor for a massless scalar.  
This is straightforward, but for general $r$ the expressions are
somewhat lengthy.  We are interested specifically in $T_{vv}(r_h)$,
since this tells us the ingoing flux of energy into the horizon.
Looking at eqn. \ref{tvv}, we see that 
\begin{equation}
T_{vv}(r_h)={1\over{192\pi}}\left[
\left(\left.{{df}\over{dr}}\right|_{r=r_c}\right)^2-
\left(\left.{{df}\over{dr}}\right|_{r=r_h}\right)^2\right]
=\sigma_{2D}\left(T_{\rm GH}^2-T_H^2\right)
\label{eq-tvvexact}
\end{equation}

The surface gravity takes the form $\kappa=GM/r_H^2 - \Lambda r_H/3$;
inserting the expressions for $r_c,r_h$ and dividing by $2\pi$ gives
$T_{\rm GH},T_H$. 
The exact expressions are lengthy, but to leading order in 
$GM\sqrt{\Lambda}$ the horizon temperatures become
\begin{eqnarray}
T_H&=& {1\over{8\pi GM}}\left[1-{{16G^2M^2\Lambda}\over 3}+
{\cal O}(G^4M^4\Lambda^2)\right]\\
T_{\rm GH}&=&{1\over{2\pi}}\sqrt{\Lambda\over 3}\left[1-2GM
\sqrt{\Lambda\over 3}+{\cal O}(M^2\Lambda)\right]
\end{eqnarray}
Since $GM\sqrt{\Lambda}\approx 1.6\ten{-23}(M/\msun) \ll 1$ for all known
black holes, the correction to the ordinary temperatures are
negligible.  Clearly, all known black holes will evaporate.

With equation~[\ref{eq-tvvexact}] 
we may also find the critical mass above which
the black hole will accrete energy overall, and therefore never
evaporate.  This occurs when $T_{vv}(r_h)=0$, implying that 
the two horizon temperatures are equal.  
Evaluating the two horizon surface gravities and equating them
gives $r_c=r_h$; that is, the critical mass is that for which the
black hole radius equals the cosmological radius.
This occurs at $M_{\rm crit}=1/(3G\sqrt{\Lambda}) \approx 2.1 \ten{22}\msun$,
close to the result from the naive estimate above.  
However, for this value of
$M$, $f=0$ and the time coordinate $t$ is nowhere timelike, 
so our coordinate system is not valid and our analysis breaks down. 
See Bousso and Hawking \cite{bousso} for a detailed discussion of black
holes with masses near and at $M_{\rm crit}$.
In any case, for all $M<M_{\rm crit}$ our analysis
should be valid, showing that the Hawking
radiation exceeds the Gibbons-Hawking accretion.  We may safely conclude
that black holes in the observed mass range evaporate.

\section{Adams, Mbonye, and Laughlin calculation}
Adams, Mbonye, \& Laughlin\cite{aml} discuss the Gibbons-Hawking correction
to the Hawking radiation and state that it is orders of magnitude
too small to affect normal black hole evolution, just as we concluded
above.  However, to take account
of the cosmological constant, they rely on a conjecture that
\begin{equation}
\label{eq-mallettmdot}
{\dot M}_H  \approx 4\pi \sigma r_H ^2 (T_{\rm Vac}^4 - T_H^4),
\end{equation}
where the effective temperature $T_{\rm Vac}$ is found from setting 
$\rho_V = a T_{\rm Vac}^4$, with $a= 4 \sigma = \pi^2/15$.
Thus $T_{\rm Vac} = 30 (\Omega_V/0.7){\rm K}$, and the corresponding
critical mass black hole from $\dot M=0$ is $M_{crit}= 2.1 \ten{-9}\msun$.
This critical black hole size implies that all known black holes are
accreting vacuum energy at a rate larger than they are emitting
Hawking radiating, and so will never evaporate.

The conjecture of Adams, Mbonye, and Laughlin is based upon a paper by
Mallett \cite{malletta} which considered black hole evaporation during
GUT scale inflation, and in particular his equation 4.8, (which is
same as as equation~[\ref{eq-mallettmdot}] with $T_{\rm Vac}$ replaced
by $T_{GUT}$).  Clearly, since $T_{\rm Vac}\neq T_{\rm GH}$, this
equation conflicts with our formulae above, and we argue that
Mallett's equation 4.8 is in error.  Our argument is strengthened by
the fact that Mallett \cite{mallettb} derives an expression for
$T_{vv}$ which, besides the error mentioned in the previous section,
has the same form as ours in the relevant limit, and so gives nearly
the same critical mass as we find above.  Thus Mallett's equation 4.8
seems to be in conflict with his calculation of $T_{vv}$.

\section{Discussion and Conclusions}
We find that the energy absorbed by black holes in a de Sitter
Universe is characterized by the very low Gibbons-Hawking temperature
and not the equivalent temperature of a radiation field with a density
of the vacuum energy as suggested by Mallett \cite{malletta}.
Adams, Mbonye, and Laughlin conjecture that the cosmological constant
is caused by a sea of virtual particles, and that black holes may
swallow those virtual particles and thereby accrete energy.  
An argument against this idea comes from the difference
between the vacuum expectation value (VEV) of a field, 
$\langle\phi\rangle$, and its potential
$V(\langle\phi\rangle)$.  Adams et al. note that the sea of virtual
particles ``must contribute a net positive energy density'', but this fact
is actually irrelevant.  What matters for $\Lambda$ is not just the
vacuum energy, but the renormalized vacuum energy, which need not be
positive despite the nonzero VEV.  Indeed, even if $\Lambda=0$ there
still exist fields with nonzero VEV, e.g. the Higgs condensate.
Since Adams et al. accept that their conjectured process would not
occur for $\Lambda=0$, and since shifting the zero point of the
potential will shift $\Lambda$ without changing the VEV, 
we see no reason that an 
``extra" accretion of particles should exist for nonzero $\Lambda$.

Besides the above, there is another argument against this conjecture.
Adams et al. 
attribute the unexpected increase in black hole mass in
their model to swallowing of virtual particles.  Accepting for
argument's sake this heuristic description, we note that in order to
have any effect on the black hole mass, these virtual particles must
go on mass shell.  Within the context of this heuristic picture, this
``on-shell" requirement is why black holes can Hawking radiate.  One
member of a virtual pair can go on shell inside the horizon, and carry
with it negative energy as viewed from infinity, while its partner can
go on shell with positive energy and escape to infinity.  In the
ordinary (no $\Lambda$) Hawking process, the black hole cannot gain
mass by such processes; the negative energy particle can't propagate
outside the horizon, and if the positive energy particle joins its
partner and is swallowed, nothing has changed.  Adams et al.
conjecture that a cosmological constant alters this process in such a
way that a thermal bath of particles at temperature (viewed from
infinity) of $\sim 30$K can go down the hole.  Note that there is no
restriction upon positive energy particles; they can exist equally
well far from the hole as near the horizon.  So if there were 30K
particles going down the hole, a 30 K thermal bath would also be seen
elsewhere (e.g. near the Earth), in contradiction with experiment.

In summary, there are several arguments against the heuristic picture
that gave rise to the conjecture that black holes accrete vacuum energy,
and our calculations above show that the only accretion is
of the Gibbons-Hawking radiation.  Finally, several
previous authors such as 
Bousso and Hawking\cite{bousso} and Davies, 
Ford, and Page\cite{fordb}
reach conclusions about black hole evaporation in
concordance with ours.  
We conclude that in a Universe dominated by vacuum energy,
all isolated black holes will eventually evaporate.

\noindent
{\bf Acknowledgements \hfil}

We would like to thank Larry Ford, Dan Holz, Bob Sanders, and Art Wolfe
for helpful discussions.
This work was supported in part by the U.S. Department of Energy
under grant DEF03-90-ER 40546.

\end{document}